\documentclass[amsmath,amssymb,nofootinbib]{revtex4}

\usepackage{latexsym}
\usepackage{amssymb}
\usepackage{amsfonts}
\usepackage{amsmath}
\usepackage[dvips]{graphicx}
\usepackage{bm}

\def\ber{\begin{eqnarray}}
\def\eer{\end{eqnarray}}
\def\bf{\begin{framed}}
\def\ef{\end{framed}}
\def\beq{\begin{equation}}
\def\eeq{\end{equation}}

\begin{document}

\title{Mapping Cartesian Coordinates into Emission Coordinates: some Toy Models}

\author{Matteo Luca Ruggiero}
\email{matteo.ruggiero@polito.it}
\author{Angelo Tartaglia}
\email{angelo.tartaglia@polito.it} \affiliation{Dipartimento di Fisica, Politecnico di Torino, Corso Duca degli
Abruzzi, 24, I-10129 Torino, Italy \\ INFN - Sezione di Torino, Via P. Giuria 1, I-10125 Torino, Italy }

\date{\today}

\begin{abstract}
After briefly reviewing the relativistic approach to positioning systems based on the introduction of the
emission coordinates, we show how explicit maps can be obtained between the Cartesian coordinates and the
emission coordinates, for suitably chosen set of emitters, whose world-lines are supposed to be known by the
users. We consider Minkowski space-time and the space-time where a small inhomogeineity is introduced (i.e. a
small "gravitational" field), both in 1+1 and 1+3 dimensions.
\end{abstract}

\maketitle


\section{Introduction}


The current positioning systems, such as GPS, are based on a classical space and an absolute time, over which
some relativistic corrections are added, in order to take into account the effects arising from both the Special
and General Theory of Relativity (see \cite{ashby1,ashby2} and references therein). A new approach to the
problem of positioning has been proposed by B. Coll\cite{collsypor}, which introduces a shift from the Newtonian
viewpoint to a true relativistic framework, where a new and operational definition of space-time coordinates is
given.

The starting assumption in the construction of such a new system of coordinates is that an ideal electromagnetic
signal propagates along a null geodesic. Indeed, the main idea can be summarized as follows: let us consider 4
clocks, moving along arbitrary world-lines in space-time, and broadcasting their proper times, by means of
electromagnetic signals. Then, any observer, at a given space-time point $P$ along his own world-line, receives
4 numbers, carried by the 4 signals emitted by the clocks. These 4 numbers, say $\tau_1,\tau_2,\tau_3,\tau_4$,
are nothing but the proper times of the emitting clocks and constitute \textit{the coordinates of that
space-time point}; they are usually referred to as \textit{emission coordinates} \cite{collem,ferrando}. In
other words, the past light cone of a space-time point $P$ cuts the clocks world-lines at 4 points, and the
proper times measured along the clocks world-lines, are the coordinates of $P $. In practice, the clocks are
supposed to be carried by satellites (which, in what follows, are referred to also as \textit{emitters})
orbiting the Earth, and the observers are the \textit{users} on the Earth, or on board of other satellites also.

Actually, a 2-dimensional approach to this new paradigm of relativistic
positioning has been thoroughly described by Coll and collaborators \cite%
{coll1a,coll2a}: there, the definition of the \textit{coordinates domain} (i.e. the space-time region where
emission coordinates are well defined), the information that the data coming from a relativistic positioning
system give on the space-time metric interval and the interest of these results in gravimetry, are discussed and
analyzed for some prototypical situations.

In this paper, starting from the results obtained in \cite{coll1a,coll2a}, we show how an explicit map can be
obtained from the Cartesian coordinates and the emission coordinates, for suitably chosen set of emitters, whose
world-lines are supposed to be known by the users. We start from the 2-dimensional cases, both in inertial and
accelerated frames, and show how the results can be generalized to 4-dimensional cases, in order to deal with
Minkowski 1+3 dimensional space-time and the space-time where a small inhomogeneity is introduced (i.e. a small
"gravitational" field).


\section{Emission coordinates in 1+1 dimensional space-time}

\label{sec:em1+1} 

In this Section, we briefly review how emission coordinates can be introduced in a 1+1 dimensional space-time,
according to the approach outlined in \cite{coll1a}.

\begin{figure}[tp]
\begin{center}
\includegraphics[width=5cm,height=5cm]{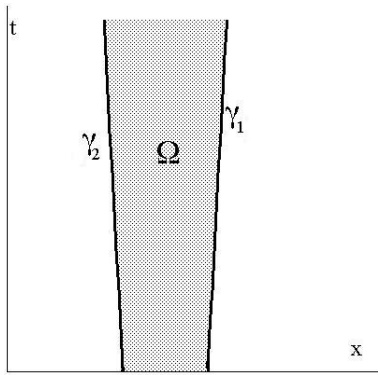}
\end{center}
\caption{The two emitters move along the world-lines $\protect\gamma_1,%
\protect\gamma_2$, and every event in the domain $\Omega$ between both
emitters can be unambiguously labelled by the proper times $(\protect\tau^1,%
\protect\tau^2)$.} \label{fig:xt}
\end{figure}

The simplified system we are dealing with is composed of (i) two emitters, whose world-lines are $\gamma _{1}$,
$\gamma _{2}$; they broadcast to the users their proper times $\tau ^{1},$ $\tau ^{2}$ and, also, the proper
times $\overline{\tau }^{2}$, $\overline{\tau }^{1}$ that they receive each one from the other; (ii) a generic
user, whose world-line is $\gamma $. The user is supposed to be in $\Omega $, which is the region of space-time
where the emission coordinates are properly defined; in the situation we are considering, $\Omega $, which is
called coordinate domain, is the
space-time region between the world-lines of the two emitters (see figure %
\ref{fig:xt}). The user, is supposed to receive the four broadcasted times $%
\{\tau ^{1},\tau ^{2};\overline{\tau }^{1},\overline{\tau }^{2}\}$. We recall that the set of these proper times
allows the user to determine the equation of his trajectory $\tau ^{2}=F(\tau ^{1}),$ and, also, the
trajectories of the emitters $\varphi _{1}(\tau ^{1})=\overline{\tau }^{2}$, $\varphi _{2}(\tau
^{2})=\overline{\tau }^{1}$.

Now, we suppose to know the world-lines of the emitters in a suitable system of null coordinates $\{u,v\}$. We
remember (see, for instance \cite{coll1a})
that, in such a coordinate system, the space-time metric has the form%
\footnote{%
The space-time metric has signature $(1,-1,-1,-1)$, and we use units such that c=1.}.

\begin{equation}
ds^{2}=m( u,v)dudv .  \label{eq:ds2nulluv1}
\end{equation}

The world-lines of the emitters, in terms of their proper times, are given by the following expressions:
\begin{equation}  \label{eq:gamma12uv}
\gamma_1 \equiv
\begin{cases}
{u} = u_1(\tau^1) \\
{v} = v_1(\tau^1)%
\end{cases}
\qquad \gamma_2 \equiv
\begin{cases}
{u} = u_2(\tau^2) \\
{v} = v_2(\tau^2)%
\end{cases}%
.
\end{equation}

Then, the emission coordinates $\{\tau^1,\tau^2\}$ are defined by the following change of variables from the
null coordinates $\{{u},{v}\}$:
\begin{equation}  \label{eq:defemcoorduvt1t21}
\begin{array}{l}
{u} = u_1(\tau^1) \\
{v} = v_2(\tau^2)%
\end{array}
\Rightarrow
\begin{array}{l}
\tau^1 = u_1^{-1}({u}) = \tau^1({u}) \\
\tau^2 = v_2^{-1}({v}) = \tau^2({v}).%
\end{array}%
\end{equation}

Consequently, the emitter world-lines (\ref{eq:gamma12uv}) can be expressed in emission coordinates, and they
have the following expression:

\begin{equation}  \label{eq:trajdeftau1tau2}
\gamma_1 \equiv
\begin{cases}
\tau^1 = \tau^1 \\
\tau^2 = \varphi_1(\tau^1) \doteq \overline{\tau}^2%
\end{cases}
\qquad \gamma_2 \equiv
\begin{cases}
\tau^1 = \varphi_2(\tau^2) \doteq \overline{\tau}^1 \\
\tau^2 = \tau^2%
\end{cases}%
\end{equation}
where the functions $\varphi_1(\tau^1)$, $\varphi_2(\tau^2)$ are determined on considering that the events along
the two world-lines are connected by coordinate lines $v=const$ and $u=const$ (see below). We remember also
that, in terms of the emission coordinates, we may write the metric in the form (thanks to the change of
coordinates (\ref{eq:defemcoorduvt1t21})):

\begin{equation}
\begin{array}{c}
ds^{2}=m(\tau ^{1},\tau ^{2})d\tau ^{1}d\tau ^{2} \\[4mm]
m(\tau ^{1},\tau ^{2})={m}(u_{1}(\tau ^{1}),v_{2}(\tau ^{2}))u_{1}^{\prime
}(\tau ^{1})v_{2}^{\prime }(\tau ^{2}).%
\end{array}
\label{eq:metricecdef1}
\end{equation}


\section{Emission coordinates for inertial emitters in Minkowski space-time}

\label{ssec:2dflatin} 

\begin{figure}[tp]
\begin{center}
\includegraphics[width=5cm,height=5cm]{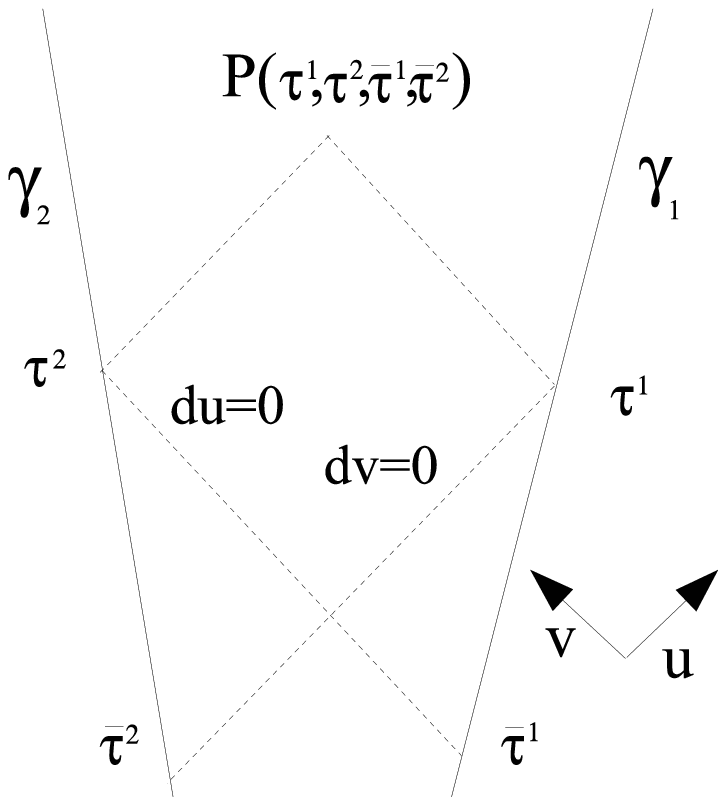} %
\includegraphics[width=5cm,height=5cm]{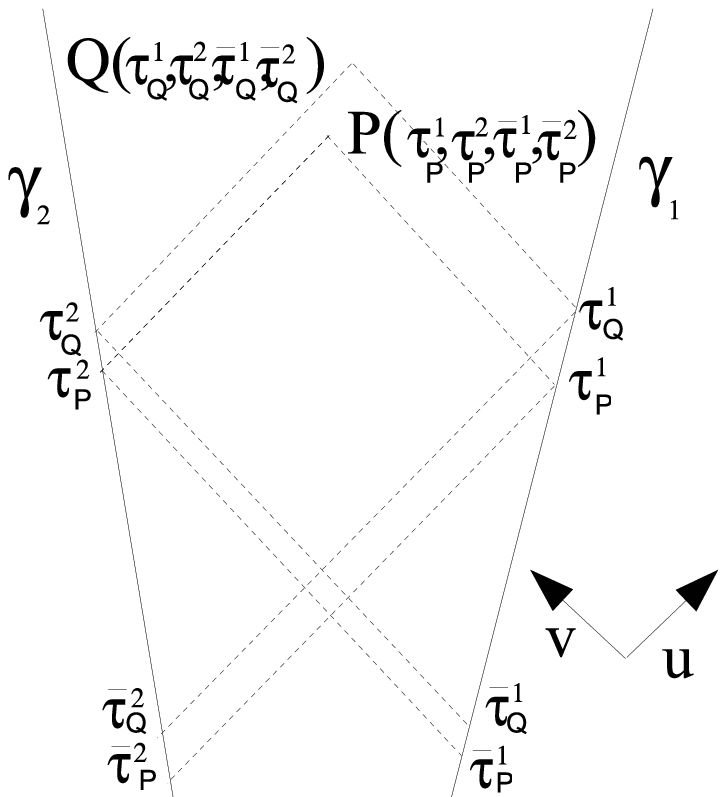}
\end{center}
\caption{ (a) The events corresponding to proper times $\protect\tau^1$ along the world-line $\protect\gamma_1$
and $\overline{\protect\tau}^2$
along the world-line $\protect\gamma_2$ are connected by a coordinate line $%
v=const$ and, hence, they have the same coordinate $v$; similarly, the
events corresponding to proper times $\protect\tau^2$ along the world-line $%
\protect\gamma_2$ and $\overline{\protect\tau}^1$ along the world-line $%
\protect\gamma_1$ are connected by a coordinate line $u=const$ and, hence, they have the same coordinate $u$.
(b) The positioning data received at $P$ by the user, thanks to the properties of the coordinate lines, allow to
establish that $u_1(\overline{\protect\tau}^1_P)=u_2({\protect\tau}^2_P)$, $%
v_1({\protect\tau}^1_P)=v_2(\overline{\protect\tau}^2_P)$, and similarly for
positioning data received at $Q$: $u_1(\overline{\protect\tau}^1_Q)=u_2({%
\protect\tau}^2_Q)$, $v_1({\protect\tau}^1_Q)=v_2(\overline{\protect\tau}%
^2_Q)$ } \label{fig:2d_12}
\end{figure}

As an example of what we have outlined above, we show how to build emission coordinates for a set of two
emitters, moving with constant velocity $v_i,\ i=1,2$ in Minkowski 1+1 dimensional space-time.

To begin with, we write their world-lines in null coordinates, as functions of the proper times $\tau^1,\tau^2$:
\begin{equation}  \label{eq:flatintraj1}
\gamma_1 \equiv
\begin{cases}
\, {u} = \lambda_1 \tau^1 \\[1mm]
\, {v} = \displaystyle \frac{1}{\lambda_1} \tau^1 + v_0%
\end{cases}
\ \gamma_2 \equiv
\begin{cases}
\, {u} = \lambda_2 \tau^2 + u_0 \\[1mm]
\, {v} =\displaystyle \frac{1}{\lambda_2} \tau^2%
\end{cases}%
,
\end{equation}

where $\lambda_i \doteq \sqrt{\frac{1+v_i}{1-v_i}}, \quad i=1,2$ are the \textit{shift functions} of the two
emitters (see \cite{coll1a}).

Then, the emission coordinates are defined by:
\begin{equation}  \label{coordinatechangeinertial}
\begin{array}{lll}
{u} = u_1(\tau^1) = \lambda_1 \tau^1 & \Rightarrow & \tau^1 = \displaystyle
\frac{1}{\lambda_1} \, {u} \\[2mm]
{v} = v_2(\tau^2) = \displaystyle \frac{1}{\lambda_2} \tau^2 & \Rightarrow &
\tau^2 = \lambda_2 {v}.%
\end{array}%
\end{equation}

Furthermore, on applying (\ref{eq:metricecdef1}), we can easily obtain the
metric tensor in emission coordinates (in the case of a straight line it is $%
m\left( u,v\right) =1$):

\begin{equation}
m(\tau ^{1},\tau ^{2})=u_{1}^{\prime }(\tau ^{1})v_{2}^{\prime }(\tau ^{2})=%
\frac{\lambda _{1}}{\lambda _{2}},  \label{eq:defflat2dmetric1}
\end{equation}

and the space-interval turns out to be

\begin{equation}
ds^2 = \lambda \, d \tau^1 d \tau^2 \, , \qquad \lambda \equiv \frac{%
\lambda_1}{\lambda_2} .  \label{eq:defflat2dmetric2}
\end{equation}

Now, let us work out how to express the emitters world-lines in emission coordinates. As we wrote before, this
can be done if we admit that the
emitters broadcast their proper times $\overline{\tau }^{2}$, $\overline{%
\tau }^{1}$ to each other. Then, the functions $\varphi _{1}(\tau ^{1})$, $%
\varphi _{2}(\tau ^{2})$ are determined by considering that the events corresponding to proper times $\tau ^{1}$
along the world-line $\gamma _{1}$ and $\overline{\tau }^{2}$ along the world-line $\gamma _{2}$ are connected
by a coordinate line $v=const$, similarly for the the events corresponding
to proper times $\tau ^{2}$ along the world-line $\gamma _{2}$ and $%
\overline{\tau }^{1}$ along the world-line $\gamma _{1}$ (see figure \ref%
{fig:2d_12}a).

Consequently, the emitters world-lines in emission coordinates are

\begin{eqnarray}  \label{emit-taus-inertial-2}
\gamma_1 \equiv
\begin{cases}
\tau^1 = \tau^1 \\
\tau^2 = \varphi_1(\tau^1) \equiv \displaystyle \frac{1}{\lambda} \tau^1 +
\tau^2_0%
\end{cases}
\\[1mm]
\gamma_2 \equiv
\begin{cases}
\tau^1 = \varphi_2(\tau^2) \equiv \displaystyle \frac{1}{\lambda} \tau^2 +
\tau^1_0 \\
\tau^2 = \tau^2.%
\end{cases}%
\end{eqnarray}

Let us now consider the set of positioning data $\left({\tau}^1_P,{\tau}^2_P,%
\overline{\tau}^1_P,\overline{\tau}^2_P\right)$, $\left({\tau}^1_Q,{\tau}%
^2_Q,\overline{\tau}^1_Q,\overline{\tau}^2_Q\right)$ received by the user at the events $P$ and $Q$,
respectively. Due to the fact that the corresponding emission events are connected by coordinate lines $u=const$
or $v=const$, it is possible to show that

\begin{equation}
\frac{\Delta \tau^1 \Delta \tau^2}{\Delta \overline{\tau}^1 \Delta \overline{%
\tau}^2} = \frac{\lambda_1^2}{\lambda_2^2}=\lambda^2, \label{eq:deltataulambda1}
\end{equation}

where $\Delta \tau^1 \doteq \tau^1_Q-\tau^1_P, \quad \Delta \tau^2 \doteq
\tau^2_Q-\tau^2_P \quad$, $\Delta \overline{\tau}^1 \doteq \overline{\tau}%
^1_Q-\overline{\tau}^1_P, \quad \Delta \overline{\tau}^2 \doteq \overline{%
\tau}^2_Q-\overline{\tau}^2_P$.

Consequently, thanks to (\ref{eq:defflat2dmetric2}), in terms of the positioning data $\{\tau^1, \tau^2;
\overline{\tau}^1, \overline{\tau}^2\}$, the space-time metric is given by
\begin{equation}  \label{general}
ds^2 = \sqrt{\frac{\Delta\tau^1 \Delta\tau^2}{\Delta\overline{\tau}^1 \Delta%
\overline{\tau}^2}} \, d \tau^1 d \tau^2
\end{equation}

\begin{figure}[top]
\begin{center}
\includegraphics[width=5cm,height=5cm]{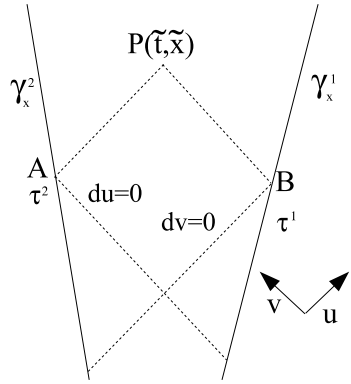}
\end{center}
\caption{Since signals propagate along the world-lines $u=const$ or $v=const$%
, there are the following relations among the coordinates of the point $P$ and the emission points $A,B$: $v(P)
= v(A)$, $u(P) = u(B)$.} \label{fig:4d_1}
\end{figure}

Now, let us show how to write a map between the usual cartesian coordinates $%
x,t$ and the emission coordinates $\tau^1,\tau^2$. To fix the ideas, we consider two satellites moving along the
$x$ axis of a Cartesian system of
coordinates. Let $P$ be an event, whose coordinates are $\tilde{t},\tilde{x}$%
: these coordinates can be expressed in terms of emission coordinates which, we recall, are the proper times
measured along the satellites world-lines.
It is useful to remember the expression of the world-lines of the emitters (%
\ref{eq:flatintraj1}):

\begin{equation}  \label{eq:4dfgammax1}
\gamma^1 \equiv
\begin{cases}
\, {u} = \lambda^1 \tau^1 \\[1mm]
\, {v} = \displaystyle \frac{1}{\lambda^1} \tau^1 + v_0%
\end{cases}
\ \gamma_2 \equiv
\begin{cases}
\, {u} = \lambda^2 \tau^2 + u_0 \\[1mm]
\, {v} =\displaystyle \frac{1}{\lambda^2} \tau^2.%
\end{cases}%
\end{equation}

Furthermore, the null coordinates $u,v$ have the following expression, in terms of the Cartesian ones:

\begin{equation}
{u}= {t}+ {x} \quad \quad {v}= {t}- {x}.  \label{eq:defuvxt}
\end{equation}

Consequently, the null coordinates of the event $P$ are

\begin{equation}
\tilde{u}=\tilde{t}+\tilde{x} \quad \quad \tilde{v}=\tilde{t}-\tilde{x}. \label{eq:4dftx2uv1}
\end{equation}

Since signals propagate along the world-lines $u=const$ or $v=const$, we may write the following relations among
the coordinates of the point $P$ and the emission points $A,B$ (see figure \ref{fig:4d_1})
\begin{eqnarray}
v(P) & = & v(A)  \label{eq:4dfvPA1} \\
u(P) & = & u(B)  \label{eq:4dfvPB1}
\end{eqnarray}

Eqs. (\ref{eq:4dfvPA1}),(\ref{eq:4dfvPB1}), together with (\ref%
{eq:4dfgammax1}) and (\ref{eq:4dftx2uv1}) imply

\begin{eqnarray}
\tilde{t}-\tilde{x} & = & \frac{\tau^2}{\lambda^2}  \label{eq:4dftxtau21} \\
\tilde{t}+\tilde{x} & = & \lambda^1 \tau^1  \label{eq:4dftxtau11}
\end{eqnarray}

from which the relations between the cartesian coordinates $\tilde{t},\tilde{%
x}$ and emission coordinates $\tau^1,\tau^2$ of the point $P$ is established:

\begin{equation}
\tilde{t}= \frac{1}{2} \left(\lambda^1 \tau^1+\frac{\tau^2}{\lambda^2}
\right) \quad \quad \tilde{x}= \frac{1}{2} \left(\lambda^1 \tau^1-\frac{%
\tau^2}{\lambda^2} \right)  \label{eq:4dftau12tx1}
\end{equation}


\section{Emission coordinates for stationary emitters in accelerated
space-time}

\label{ssec:accst} 

Another example of definition of emission coordinates in a 1+1 dimensional space-time that can be dealt with in
full details is that of stationary emitters in an accelerated space-time. The metric of an accelerated
space-time is give by (see \cite{moller})

\begin{equation}
ds^2= (1+gx)^2dt^2-dx^2,  \label{eq:accmetric1}
\end{equation}
where $g=constant$.

Now, let us apply the coordinate change

\begin{equation}
X=\frac{1}{g} \left(\cosh g t-1 \right)+ x\cosh g t \quad T=\frac{1}{g}%
\left(\sinh g t \right)+x \sinh gt,  \label{eq:accXTxt}
\end{equation}

so that the metric (\ref{eq:accmetric1}) becomes

\begin{equation}
ds^2=dT^2-dX^2 .  \label{eq:accmetricXT}
\end{equation}

That being done, we can easily introduce the null coordinates

\begin{equation}
U=T+X \quad V=T-X ,  \label{eq:accTXUV1}
\end{equation}

and, consequently, the metric assumes the form

\begin{equation}
ds^2=dUdV ,  \label{eq:accmetricUV}
\end{equation}

which easily allows us to define emission coordinates, for a suitable class of emitters. Namely, let us consider
\textit{stationary emitters}, i.e. emitters that are at rest in the metric (\ref{eq:accmetric1}): in other
words, their world-lines are

\begin{equation}
x=x^i, \quad i=1,2,  \label{eq:accwl1}
\end{equation}

where $x^{i}$ are constant. By means of (\ref{eq:accmetric1}),(\ref%
{eq:accXTxt}) we may write the world-lines, in terms of proper time, in the form

\begin{equation}
\gamma _{i}\equiv
\begin{cases}
\,X^{i}=\frac{1}{g}\left[ \cosh \left( \frac{g(\tau ^{i}-\tau _{0}^{i})}{%
1+gx^{i}}\right) -1\right] +x^{i}\cosh \left( \frac{g(\tau ^{i}-\tau
_{0}^{i})}{1+gx^{i}}\right)  \\[1mm]
\,T^{i}=\frac{1}{g}\sinh \left( \frac{g(\tau ^{i}-\tau _{0}^{i})}{1+gx^{i}}%
\right) +x^{i}\sinh \left( \frac{g(\tau ^{i}-\tau _{0}^{i})}{1+gx^{i}}%
\right) ,%
\end{cases}
\label{eq:accwlXT}
\end{equation}
Where $\tau _{0}^{i}$ defines the relation between the proper time $\tau^i$ and coordinate time $t^i$ of the
emitters, according to \beq \tau ^{i}-\tau _{0}^{i}=\left( 1+gx^{i}\right) t^{i} \label{eq:reltaut1} \eeq

Hence, thanks to (\ref{eq:accTXUV1})

\begin{equation}  \label{eq:accwlUV}
\gamma_i \equiv
\begin{cases}
\, U^i = \exp\left(\frac{g(\tau^i-\tau^i_0)}{1+g x^i} \right) \left(\frac{1}{%
g}+x^i \right)-\frac{1}{g} \\[1mm]
\, V^i =-\exp\left(-\frac{g(\tau^i-\tau^i_0)}{1+g x^i} \right)\left(\frac{1}{%
g}+x^i \right)+ \frac{1}{g}%
\end{cases}%
.
\end{equation}

The emission coordinates are defined as follows (see (\ref%
{eq:defemcoorduvt1t21})):

\begin{equation}  \label{eq:accdeftau1tau2}
\begin{array}{l}
{U} = U_1(\tau^1) =\exp\left(\frac{g(\tau^1-\tau^1_0)}{1+g x^1} \right)
\left(\frac{1}{g}+x^1 \right)-\frac{1}{g} \\[2mm]
{V} = V_2(\tau^2) = -\exp\left(-\frac{g(\tau^2-\tau^2_0)}{1+g x^2}
\right)\left(\frac{1}{g}+x^2 \right)+ \frac{1}{g}%
\end{array}%
.
\end{equation}

Then, the emitters world-lines, in emission coordinates turn out to be (see
eq. (\ref{eq:trajdeftau1tau2}) and the discussion in Section \ref%
{ssec:2dflatin}):

\begin{eqnarray}  \label{eq:accwltautau21}
\gamma_1 & \equiv &
\begin{cases}
\tau^1 = \tau^1 \\
\tau^2 = \frac{1}{k}(\tau^1 - q - \sigma) \equiv \varphi_1(\tau^1)%
\end{cases}
\\
\gamma_2 & \equiv &
\begin{cases}
\tau^1 = k \tau^2 - q + \sigma \equiv \varphi_2(\tau^2) \\
\tau^2 = \tau^2,%
\end{cases}
\label{eq:accwltautau211}
\end{eqnarray}
where
\begin{eqnarray}  \label{eq:accdefkq}
k \equiv \frac{1+gx^1}{1+gx^2} > 1 , &\ & \quad q \equiv \frac{1+gx^1}{g}
\ln \frac{1+gx^1}{1+gx^2}> 0 \\[2mm]
\sigma&\equiv&\tau^1_0 - \frac{1+gx^1}{1+gx^2} \, \tau^2_0. \quad \label{eq:accdefsigma}
\end{eqnarray}

\textbf{Remark. \ } We notice that the emitters whose world-lines are (\ref%
{eq:accwlXT}) are not geodesic. In fact, we can calculate the acceleration vector

\begin{equation}
\bm{a}(\tau) \equiv \left(a^T,a^X \right)=\left(\frac{g}{1+gx} \sinh \left(%
\frac{g(\tau-\tau_0)}{1+g x}\right),\frac{g}{1+gx} \cosh \left(\frac{%
g(\tau-\tau_0)}{1+g x} \right) \right),  \label{eq:accaccdef1}
\end{equation}

from which the \textit{acceleration scalar} \beq \alpha(\tau) \doteq \sqrt{-||\bm{a}(\tau)||^2}
\label{eq:defaccascalar1} \eeq becomes:
\begin{equation}
\alpha(\tau) =\frac{g}{1+gx}. \label{eq:accdefalpha1}
\end{equation}
This fact allows us to write the relations (\ref{eq:accdefkq}) and (\ref%
{eq:accdefsigma}) in terms of the acceleration scalars of the emitters:
\begin{eqnarray}  \label{eq:accdefkq1}
k \equiv \frac{\alpha_2}{\alpha_1} > 1 , &\ & \quad q \equiv \frac{1}{%
\alpha_1} \ln \frac{\alpha_2}{\alpha_1}> 0 \\[2mm]
\sigma&\equiv&\tau^1_0 - \frac{\alpha_2}{\alpha_1} \, \tau^2_0. \quad \label{eq:accdefsigma1}
\end{eqnarray}

This makes  a direct comparison possible with the relations obtained in \cite%
{coll2a} and, consequently, the same properties apply (in particular the possibility of determining the
parameters $k,q,\sigma $).

Furthermore, eq. (\ref{eq:accdefalpha1}) suggests the physical meaning of $g$%
: it is the acceleration of the origin of the accelerated reference frame.


\section{Emission coordinates for stationary emitters in accelerated
space-time, small $g$ case}

\label{sssec:accstsmall} 

In this Section we consider, as in the previous one, stationary emitters in
an accelerated 1+1 dimensional space-time: however we suppose here that the parameter $g$%
, together with the size and position of the considered space-time region, allows for a linearization of the
metric (\ref{eq:accmetric1}) with respect to $g$. In practice we are assuming both $gx$ and $g\tau $ to be much
smaller than $1$ which means that the approximation cannot last too much in time.

To begin with, the emitters world-lines (\ref{eq:accwlXT}), after a first-order expansion in $g$ turn out to be

\begin{equation}  \label{eq:accsgwlXT}
\gamma_i \equiv
\begin{cases}
\, X^i = x^i \\[1mm]
\, T^i = \tau^i-\tau^i_0,%
\end{cases}%
\end{equation}

and, introducing as before the null coordinates $U,V$, thanks to (\ref%
{eq:accTXUV1}), the world-lines become

\begin{equation}  \label{eq:accsgwlUV}
\gamma_i \equiv
\begin{cases}
\, U^i = \tau^i-\tau^i_0+x^i \\[1mm]
\, V^i =\tau^i-\tau^i_0-x^i.%
\end{cases}%
\end{equation}

This allows us to define the emission coordinates $\left(\tau_1,\tau_2\right) $, according to
(\ref{eq:defemcoorduvt1t21})

\begin{equation}  \label{eq:accsgdeftau1tau2}
\begin{array}{l}
{U} = U_1(\tau^1) =\tau^1-\tau^1_0+x^1 \\[2mm]
{V} = V_2(\tau^2) =\tau^2-\tau^2_0-x^2,%
\end{array}%
\end{equation}

Moreover, we can write the emitters world-lines in emission coordinates (see
eq. (\ref{eq:trajdeftau1tau2}) and the discussion in Section \ref%
{ssec:2dflatin}):

\begin{eqnarray}  \label{eq:accsgwltautau21}
\gamma_1 & \equiv &
\begin{cases}
\tau^1 = \tau^1 \\
\tau^2 = \frac{1}{k}(\tau^1 - q - \sigma) \equiv \varphi_1(\tau^1)%
\end{cases}
\\
\gamma_2 & \equiv &
\begin{cases}
\tau^1 = k \tau^2 - q + \sigma \equiv \varphi_2(\tau^2) \\
\tau^2 = \tau^2,%
\end{cases}
\label{eq:accsgwltautau211}
\end{eqnarray}
where, now

\begin{eqnarray}  \label{eq:accsgdefkq}
k \equiv 1+g\left(x^1-x^2\right) , &\ & \quad q \equiv \left(x^1-x^2
\right)\left(1+g\frac{x^1+x^2}{2} \right) \\[2mm]
\sigma&\equiv&\tau^1_0 - \left[1+g\left(x^1-x^2\right) \right] \, \tau^2_0. \quad  \label{eq:accsgdefsigma}
\end{eqnarray}

Furthermore, the space-time metric has the expression

\begin{equation}
ds^2 = \exp\left[g(\tau^1 - \tau^2 - \sigma)\right] d \tau^1 d \tau^2, \label{eq:accmetrictau1tau2}
\end{equation}%

which, taking into account the smallness of the parameter $g$, becomes:

\begin{equation}
ds^2 = \left[1+g(\tau^1 - \tau^2 - \sigma)\right] d \tau^1 d \tau^2. \label{eq:accmetrictau1tau2bis}
\end{equation}%

Now, let us consider an event having coordinates $\tilde{t},\tilde{x}$. Its null coordinates are (see
(\ref{eq:accsgdeftau1tau2}) and (\ref{eq:reltaut1}))

\begin{equation}
\tilde{U}=\tilde{t} \left(1+g\tilde{x} \right)+\tilde{x} \quad \quad \tilde{V%
}=\tilde{t} \left(1+g\tilde{x} \right)-\tilde{x} \label{sec:accmetsgdefUVbar1}
\end{equation}

Then, we may write (see also Section \ref{sec:gen2d24d} and, in particular, figure \ref{fig:4d_1}):

\begin{eqnarray}
\tau^2-\tau^2_0-x^2&=&\tilde{t} \left(1+g\tilde{x} \right)-\tilde{x}
\label{eq:accsgtaux1} \\
\tau^1-\tau^1_0-x^1&=&\tilde{t} \left(1+g\tilde{x} \right)+\tilde{x} \label{eq:accsgtaux2}
\end{eqnarray}

from which we obtain $\tilde{\tau},\tilde{x}$ in terms of $\tau ^{1},\tau ^{2}$, up to first order in
$g\tilde{x}$:

\begin{eqnarray}
\tilde{t}&=&\left(1-g\tilde{x} \right)\frac{\tau^1-\tau^1_0+\tau^2-%
\tau^2_0+x^1-x^2}{2}  \label{eq:accsgttau1tau21} \\
\tilde{x}&=&\frac{\tau^1-\tau^1_0-\left(\tau^2-\tau^2_0\right)+x^1+x^2}{2} \label{eq:accsgxtau1tau22}
\end{eqnarray}

In other words, a non linear relation between $\tilde{t},\tilde{x}$ and $%
\tau^1,\tau^2$ holds.


\section{A generalization of the 2-dimensional approach to 4-dimensions}

\label{sec:gen2d24d}

\begin{figure}[top]
\begin{center}
\includegraphics[width=4cm,height=4cm]{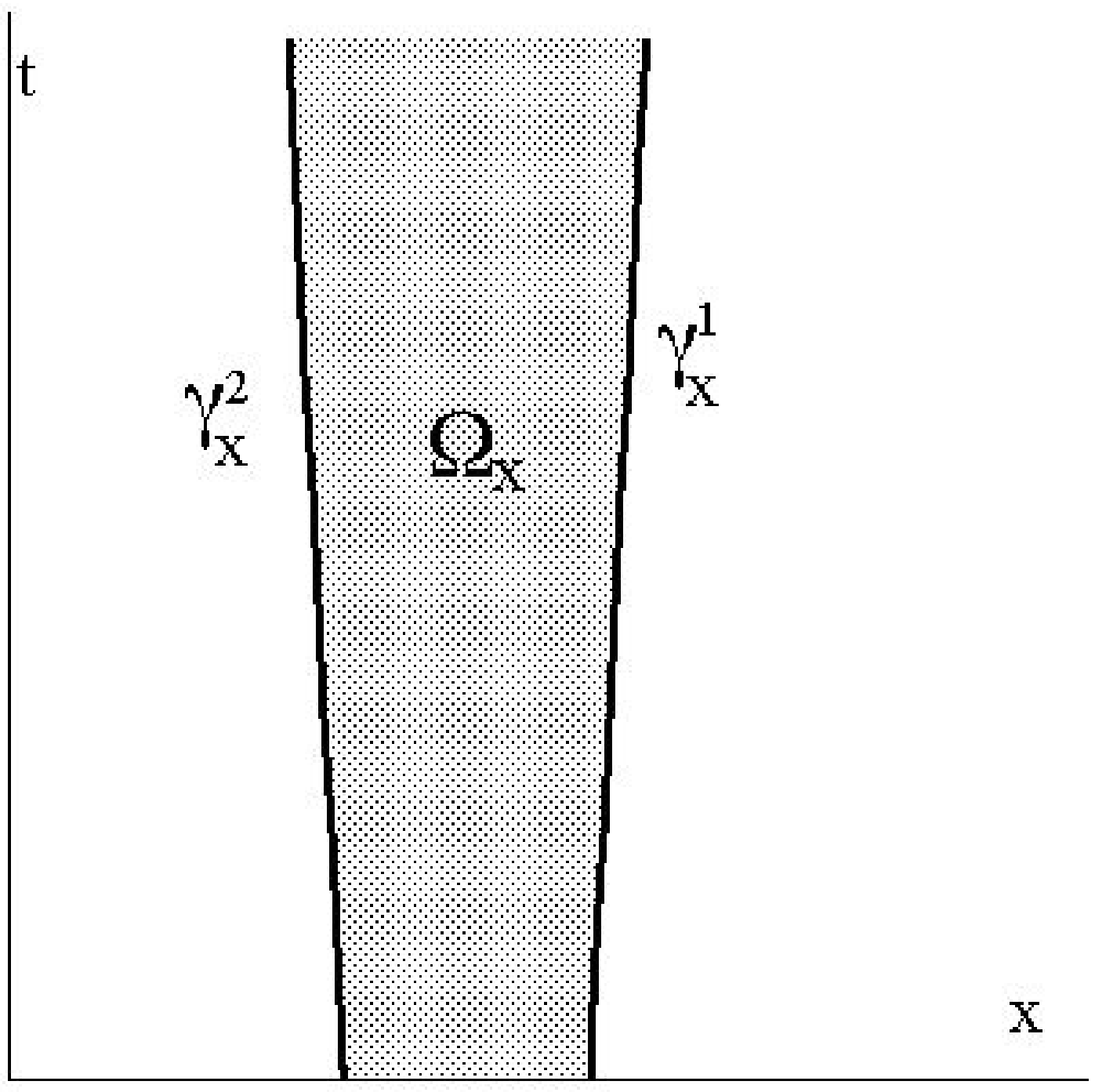} %
\includegraphics[width=4cm,height=4cm]{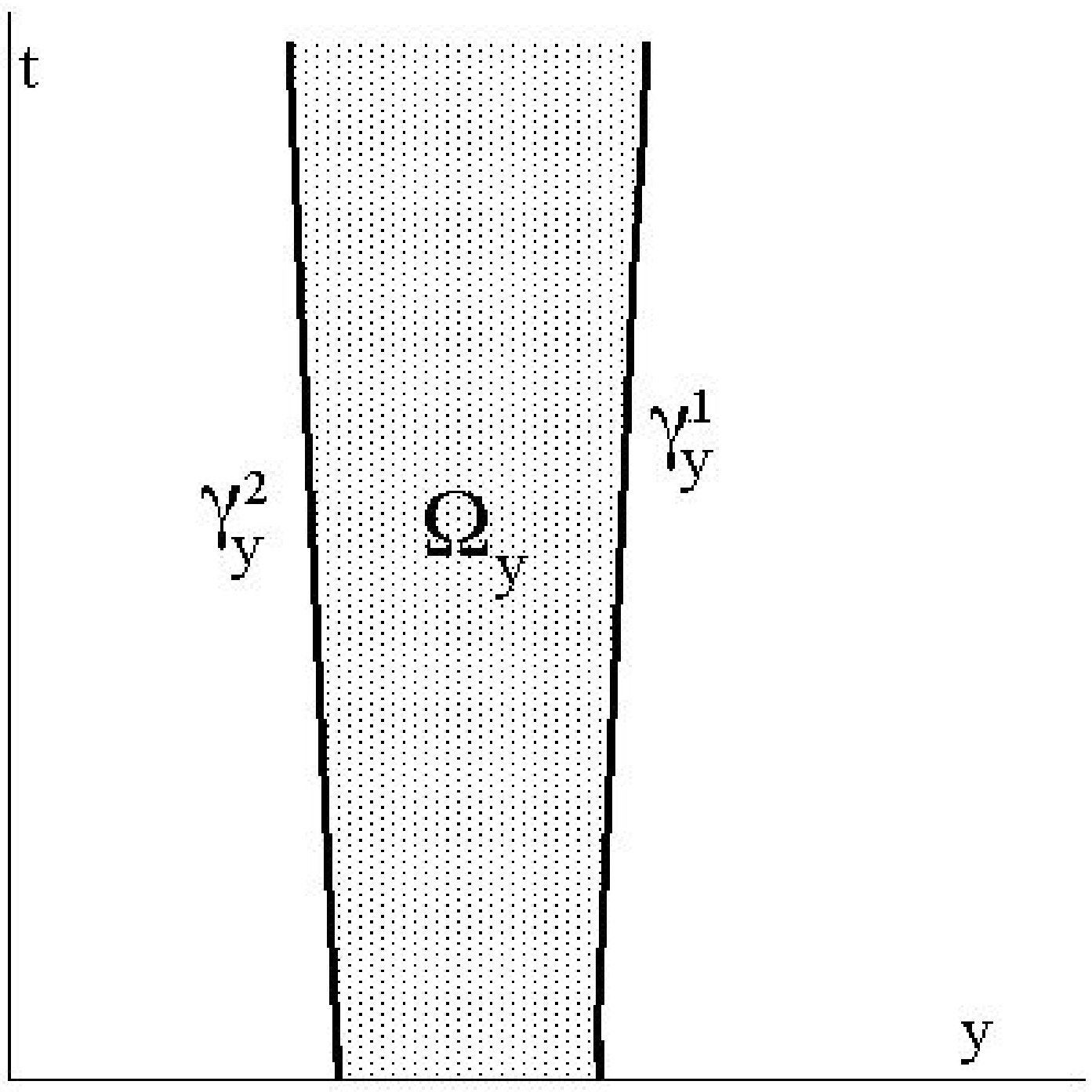} %
\includegraphics[width=5cm,height=5cm]{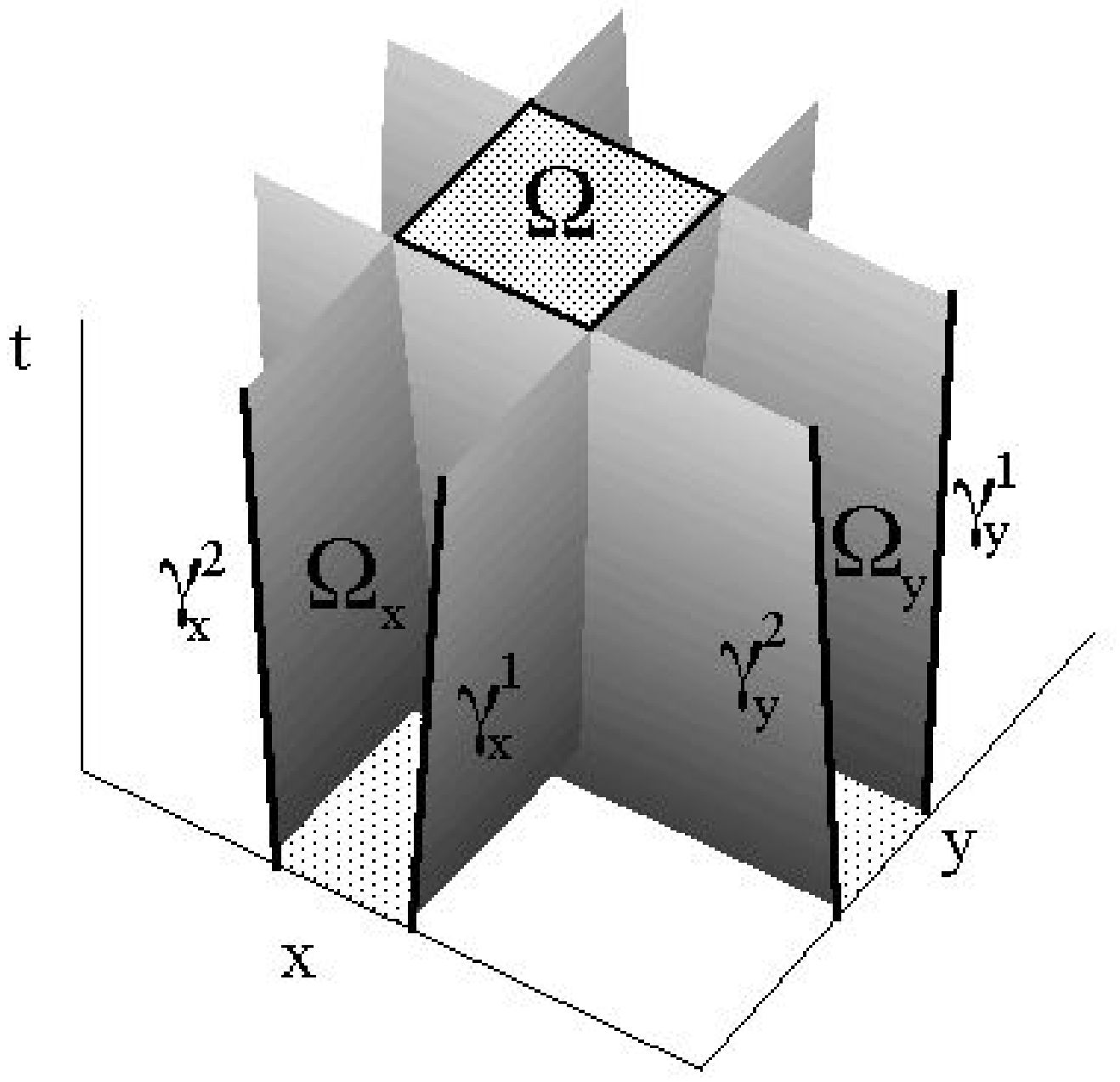}
\end{center}
\caption{The emission coordinates $\protect\tau_x^1,\protect\tau_x^2$ are defined in the coordinate domain
$\Omega_x$, which is the space-time region between the two emitters world-lines; the same holds for the emission
coordinates $\protect\tau_y^1,\protect\tau_y^2$ defined in $\Omega_y$. Then,
the set of emission coordinates $\protect\tau_x^1,\protect\tau_x^2,\protect%
\tau_y^1,\protect\tau_y^2$ is well defined in $\Omega= \Omega_x \cap \Omega_y $ } \label{fig:xyt}
\end{figure}

After having described some examples of the definition of emission coordinates in 1+1 dimensional space-times,
now we aim at exploiting the framework outlined so far in order to set up a system of emission coordinates
capable of mapping a more realistic 1+3 dimensional space-time, thanks to a suitable set of emitters. Actually,
these emitters will be chosen in order to allow an easy manage of the map between Cartesian and emission
coordinates: in other words, we will consider some toy models that, nonetheless, enable us to show the
underlying ideas of the emission coordinates approach to positioning systems.

We start, in next Section, by considering the case of a flat four dimensional space-time. Then, in the
subsequent Section, we introduce a small inhomogenity, i.e. a small gravitational field, and discuss some
possible applications. 


\subsection{Flat space-time case}\label{ssec:gen2d24dflat}

Let us consider three pairs of satellites, provided with clocks and emitters, each of them moving, with constant
speed, along a coordinate axis in a flat Minkwoski background. The particular choice of the satellites allows to
benefit from what we have done in the 2-dimensional case above. In fact, for each pair of satellites, the metric

\begin{equation}
ds^2=dt^2-dx^2-dy^2-dz^2,  \label{eq:4dfmetric1}
\end{equation}

reduces to

\begin{equation}
ds^2=dt^2-(dx^i)^2 \quad \quad i=1,2,3.  \label{eq:4dfmetric2}
\end{equation}

In particular, the metric (\ref{eq:4dfmetric2}) allows to easily introduce null coordinates $u,v$ as we have
done in Section \ref{ssec:2dflatin}: this can be done for each pair of satellites.

To be precise, recalling the notation introduced in Section \ref%
{ssec:2dflatin}, we can label the world-lines of each pair of satellites as follows

\begin{equation}  \label{eq:4dfgammaa1}
\gamma_a^1 \equiv
\begin{cases}
\, {u}_a = \lambda_a^1 \tau_a^1 \\[1mm]
\, {v}_a = \displaystyle \frac{1}{\lambda_a^1} \tau_a^1 + v_{a 0}%
\end{cases}
\ \gamma_a^2 \equiv
\begin{cases}
\, {u}_a = \lambda_a^2 \tau_a^2 + u_{a 0} \\[1mm]
\, {v}_a =\displaystyle \frac{1}{\lambda_a^2} \tau_a^2%
\end{cases}
\quad \quad a=x,y,z.
\end{equation}

Then, if we consider the pair of satellites moving along the $x$ axis, eq. (%
\ref{eq:4dftau12tx1}) tells us that the relations between the cartesian coordinates $\tilde{t},\tilde{x}$ and
emission coordinates $\tau ^{1},\tau ^{2}$ of the point $P$ are:

\begin{equation}
\tilde{t}= \frac{1}{2} \left(\lambda^1_x \tau_x^1+\frac{\tau_x^2}{\lambda_x^2%
} \right) \quad \quad \tilde{x}= \frac{1}{2} \left(\lambda_x^1 \tau_x^1-%
\frac{\tau_x^2}{\lambda_x^2} \right).  \label{eq:4dftau12tx2}
\end{equation}

The same procedure that we have just outlined can be repeated considering the other two pairs of satellites
moving along the $y$ and $z$ axes. In
doing so, we obtain the following expressions for the coordinates $\tilde{t},%
\tilde{y},\tilde{z}$ of the point $P$, in terms of the proper times $%
\tau_y^1,\tau_y^2,\tau_z^1,\tau_z^2$:

\begin{equation}
\tilde{t}= \frac{1}{2} \left(\lambda^1_y \tau_y^1+\frac{\tau_y^2}{\lambda_y^2%
} \right) \quad \quad \tilde{y}= \frac{1}{2} \left(\lambda_y^1 \tau_y^1-%
\frac{\tau_y^2}{\lambda_y^2} \right),  \label{eq:4dftau12ty2}
\end{equation}

\begin{equation}
\tilde{t}= \frac{1}{2} \left(\lambda^1_z \tau_z^1+\frac{\tau_z^2}{\lambda_z^2%
} \right) \quad \quad \tilde{z}= \frac{1}{2} \left(\lambda_z^1 \tau_z^1-%
\frac{\tau_z^2}{\lambda_z^2} \right).  \label{eq:4dftau12tz2}
\end{equation}

Summarizing, this procedure allows us to write the cartesian coordinates $%
\tilde{t},\tilde{x},\tilde{y},\tilde{z}$ of the point $P$ in terms of the 6 proper times $\tau _{x}^{1},\tau
_{x}^{2},\tau _{y}^{1},\tau _{y}^{2},\tau
_{z}^{1},\tau _{z}^{2}$. Of course, the expressions (\ref{eq:4dftau12tx2}-%
\ref{eq:4dftau12tz2}) are redundant, since the coordinate $\tilde{t}$ is the same for all. This fact can be
exploited in order to eliminate 2 of the 6 proper times, which means that 4 satellites would be enough, as can
be expected from the dimensions of space-time. For instance, we may use the constraints

\begin{equation}
\tilde{t}(\tau_x^1,\tau_x^2)=\tilde{t}(\tau_y^1,\tau_y^2), \label{eq:4dftvinc1}
\end{equation}

and

\begin{equation}
\tilde{t}(\tau_x^1,\tau_x^2)=\tilde{t}(\tau_z^1,\tau_z^2), \label{eq:4dftvinc2}
\end{equation}

to express $\tau_2^y$ as a function of $\tau_x^1,\tau_x^2,\tau_y^1$ and $%
\tau_2^z$ as a function of $\tau_x^1,\tau_x^2,\tau_z^1$, respectively.
Consequently, eqs. (\ref{eq:4dftau12tx2}-\ref{eq:4dftau12tz2}) and (\ref%
{eq:4dftvinc1}), (\ref{eq:4dftvinc2}) allows us to express the cartesian coordinates
$\tilde{t},\tilde{x},\tilde{y},\tilde{z}$ of the point $P$ in terms of the 4 proper times
$\tau_x^1,\tau_x^2,\tau_y^1,\tau_z^1$:

\begin{eqnarray}
\tilde{t} & = & \tilde{t}\left(\tau_x^1,\tau_x^2,\tau_y^1,\tau_z^1\right)
\label{eq:4dfmaptxyz2tau0t} \\
\tilde{x} & = & \tilde{x}\left(\tau_x^1,\tau_x^2,\tau_y^1,\tau_z^1\right)
\label{eq:4dfmaptxyz2tau0x} \\
\tilde{y} & = & \tilde{y}\left(\tau_x^1,\tau_x^2,\tau_y^1,\tau_z^1\right)
\label{eq:4dfmaptxyz2tau0y} \\
\tilde{z} & = & \tilde{z}\left(\tau_x^1,\tau_x^2,\tau_y^1,\tau_z^1\right). \label{eq:4dfmaptxyz2tau0z}
\end{eqnarray}

\textbf{Remark.} According to our approach, each pair of emission coordinates $\tau _{a}^{1},\tau _{a}^{2}$ is
well defined in the corresponding coordinate domain $\Omega _{a}$ (see Section \ref{sec:em1+1} and figure
\ref{fig:xt}) which, in turn, depends on the emitters world-lines. So, if we use more pairs of emission
coordinates, the full
coordinate domain $\Omega $ is the intersection of the coordinate domains $%
\Omega _{a}$. For instance, if we consider in the 1+2 dimensional space-time the emission
coordinates $\tau _{x}^{1},\tau _{x}^{2}$ defined in $\Omega _{x}$, and $%
\tau _{y}^{1},\tau _{y}^{2}$ defined in $\Omega _{y}$, the set of emission coordinates $\tau _{x}^{1},\tau
_{x}^{2},\tau _{y}^{1},\tau _{y}^{2}$ is
well defined in $\Omega =\Omega _{x}\cap \Omega _{y}$ (see figure \ref%
{fig:xyt}). This obviously can be generalized to the 1+3 dimensional space-time.\newline

If we explicitly write the equations (\ref{eq:4dfmaptxyz2tau0t}-\ref%
{eq:4dfmaptxyz2tau0z}), we see that the relation between the Cartesian and emission coordinates is given by the
following linear map:
\begin{equation}
\bm{X}= \bm{A} \cdot \bm{T},  \label{eq:4dfT2X1}
\end{equation}

where
\begin{equation}
\bm{X} \doteq \left(%
\begin{array}{c}
\tilde{t} \\
\tilde{x} \\
\tilde{y} \\
\tilde{z} \\
\end{array}%
\right),  \label{eq:4dfdefvecX1}
\end{equation}

\begin{equation}
\bm{T} \doteq \left(%
\begin{array}{c}
\tau_x^1 \\
\tau_x^2 \\
\tau_y^1 \\
\tau_z^1 \\
\end{array}%
\right),  \label{eq:4dfdefvecT1}
\end{equation}

and the coordinate change matrix $\bm{A}$ is defined by
\begin{equation}
\bm{A}\doteq \left(
\begin{array}{cccc}
\frac{\lambda _{x}^{1}}{2} & \frac{1}{2\lambda _{x}^{2}} & 0 & 0 \\
\frac{\lambda _{x}^{1}}{2} & \frac{1}{2\lambda _{x}^{2}} & 0 & 0 \\
-\frac{\lambda _{x}^{1}}{2} & -\frac{1}{2\lambda _{x}^{2}} & \lambda _{y}^{1}
& 0 \\
-\frac{\lambda _{x}^{1}}{2} & -\frac{1}{2\lambda _{x}^{2}} & 0 & \lambda
_{z}^{1}. \\
&  &  &
\end{array}%
\right)   \label{eq:4dfdefmatA1}
\end{equation}

It is easy to show that $det(\bm{A})=-\frac{\lambda _{x}^{1}\lambda _{y}^{1}\lambda _{z}^{1}}{2\lambda
_{x}^{2}}$ so that it is generally different from zero.

The inverse metric is given by

\begin{equation}
\bm{A}^{-1} \doteq \left(%
\begin{array}{cccc}
\frac{1}{\lambda^1_x} & \frac{1}{\lambda^1_x} & 0 & 0 \\
\lambda^2_x & -\lambda^2_x & 0 & 0 \\
\frac{1}{\lambda^1_y} & 0 & \frac{1}{\lambda^1_y} & 0 \\
\frac{1}{\lambda^1_z} & 0 & 0 & \frac{1}{\lambda^1_z}. \\
&  &  &
\end{array}%
\right)  \label{eq:4dfdefmatA11}
\end{equation}

Actually, the choice of the emission coordinates is somewhat arbitrary: in other words, in the approach we have
just outlined, we can choose 12 sets of emission coordinates. In fact, we must receive signals from satellites
propagating along all axes, in order to have a map of the whole space; this implies that 2 signals must come
from satellites propagating along the same direction, and the other two signals must come from satellites
propagating along the remaining directions. The latter can be arranged according to 4 combination: for instance,
if we choose that two signals come from satellites propagating along the $x$ directions, the corresponding
combinations of emission coordinates are:
\begin{equation}
\begin{array}{cccc}
\tau _{x}^{1} & \tau _{x}^{2} & \tau _{y}^{1} & \tau _{z}^{1} \\
\tau _{x}^{1} & \tau _{x}^{2} & \tau _{y}^{1} & \tau _{z}^{2} \\
\tau _{x}^{1} & \tau _{x}^{2} & \tau _{y}^{2} & \tau _{z}^{1} \\
\tau _{x}^{1} & \tau _{x}^{2} & \tau _{y}^{2} & \tau _{z}^{2} \\
&  &  &
\end{array}
\label{eq:4dfcombcoord1}
\end{equation}%
The same argument applies if we choose that two signals come from satellites propagating along the $y$ and $z$
directions, so that, summarizing, we have
12 possible choices of coordinates. We point out that the coordinate domain $%
\Omega $ is the intersection of the coordinate domain $\Omega _{a}$ (see the previous Remark). Finally, we
notice that, in this approach, the uncertainties on the Cartesian coordinates can be expressed in terms of the
uncertainties on the emission coordinates and on the parameters of the world-lines.


\subsection{Quasi-flat space-time case}\label{ssec:gen2d24dquasiflat}

Let us consider the space-time described by the metric:

\begin{equation}
ds^2=(1+gx^2)dt^2-dx^2-dy^2-dz^2.  \label{eq:4dqfmetric1}
\end{equation}

In the space-time described by (\ref{eq:4dqfmetric1}), we consider three pairs of emitters as follows: two pairs
are moving along the $y$ and $z$ axis, with constant speed, so that the metric (\ref{eq:4dqfmetric1}) reduces to

\begin{equation}
ds^2=dt^2-(dx^i)^2 \quad \quad i=2,3.  \label{eq:4dqfmetric2}
\end{equation}

The other two emitters, are at rest, at $x=x^1,x=x^2$, where $x^1,x^2$ are constant, in the $x,t$ plane (their
$y,z$ coordinates are null). For them, the metric (\ref{eq:4dqfmetric1}) reduces to

\begin{equation}
ds^{2}=(1+gx)^{2}dt^{2}.  \label{eq:4dqfmetric21}
\end{equation}

In particular, we may apply the formalism of section \ref{ssec:2dflatin} to
the first two pairs of satellites, and the formalism of section \ref%
{sssec:accstsmall} to the last pair of satellites.

On doing so, we want to express the coordinate $\tilde{t},\tilde{x},\tilde{y}%
,\tilde{z}$ of a point $P$ in terms of emission coordinates, as we have done in Section \ref{ssec:gen2d24dflat}
above.

Recalling the results of section \ref{sssec:accstsmall} (from which we borrow hypotheses and notation), we know
that we may write

\begin{equation}
\tilde{t}=\left(1-g\tilde{x} \right)\frac{\tau^1_x-\tau^1_{x0}+\tau^2_x-%
\tau^2_{x0}+x^1-x^2}{2} \quad \tilde{x}=\frac{\tau^1_x-\tau^1_{x0}-\left(%
\tau^2_x-\tau^2_{x0}\right)+x^1+x^2}{2},  \label{eq:4dqftxX}
\end{equation}

and

\begin{equation}
\tilde{t}= \frac{1}{2} \left(\lambda^1_y \tau_y^1+\frac{\tau_y^2}{\lambda_y^2%
} \right) \quad \quad \tilde{y}= \frac{1}{2} \left(\lambda_y^1 \tau_y^1-%
\frac{\tau_y^2}{\lambda_y^2} \right),  \label{eq:4dqftau12ty2}
\end{equation}

\begin{equation}
\tilde{t}= \frac{1}{2} \left(\lambda^1_z \tau_z^1+\frac{\tau_z^2}{\lambda_z^2%
} \right) \quad \quad \tilde{z}= \frac{1}{2} \left(\lambda_z^1 \tau_z^1-%
\frac{\tau_z^2}{\lambda_z^2} \right).  \label{eq:4dqftau12tz2}
\end{equation}

As a consequence, the Cartesian coordinates $\tilde{t},\tilde{x},\tilde{y},%
\tilde{z}$ of the point $P$ are known in terms of the 6 proper times $%
\tau_x^1,\tau_x^2,\tau_y^1,\tau_y^2,\tau_z^1,\tau_z^2$. Again, the expressions
(\ref{eq:4dqftxX}-\ref{eq:4dqftau12tz2}) are redundant, since the coordinate $\tilde{t}$ is the same for all.
This fact can be exploited in order to eliminate 2 of the 6 proper times. If we use the constraints

\begin{equation}
\tilde{t}(\tau_x^1,\tau_x^2)=\tilde{t}(\tau_y^1,\tau_y^2), \label{eq:4dqftvinc1}
\end{equation}

and

\begin{equation}
\tilde{t}(\tau_x^1,\tau_x^2)=\tilde{t}(\tau_z^1,\tau_z^2), \label{eq:4dqftvinc2}
\end{equation}

we may express $\tau _{2}^{y}$ as a function of $\tau _{x}^{1},\tau _{x}^{2},\tau _{y}^{1}$ and $\tau _{2}^{z}$
as a function of $\tau
_{x}^{1},\tau _{x}^{2},\tau _{z}^{1}$, respectively. Consequently, eqs. (\ref%
{eq:4dqftxX}-\ref{eq:4dqftau12tz2}) and (\ref{eq:4dqftvinc1}-\ref%
{eq:4dqftvinc2}) allow us to express the Cartesian coordinates $\tilde{t},%
\tilde{x},\tilde{y},\tilde{z}$ of the point $P$ in terms of the 4 proper times $\tau _{x}^{1},\tau _{x}^{2},\tau
_{y}^{1},\tau _{z}^{1}$. Also in this case, the coordinate domain $\Omega $ is determined by the intersection of
the coordinate domains $\Omega _{a}$ where each pair of emission coordinates $\tau _{a}^{1},\tau _{a}^{2}$ is
defined (see the Remark in Section \ref{ssec:gen2d24dflat}),

An important difference with the purely flat case described in Section \ref%
{ssec:gen2d24dflat} is that now a non linear relation between the Cartesian
and emission coordinates arises, due to the presence of the acceleration $g$%
. However, we may write the explicit relations:

\begin{eqnarray}
\tilde{t} & = & a\tau^1_x+b(\tau^1_x)^2+c\tau^2_x+d(\tau^2_x)^2+e+f+g
\label{eq:4dqft1} \\
\tilde{x} & = & \frac{\tau^1_x-\tau^1_{x0}}{2}-\frac{\tau^2_x-\tau^2_{x0}}{2}%
+\frac{x^1+x^2}{2}  \label{eq:4dqfx1} \\
\tilde{y} & = & -a\tau^1_x-b(\tau^1_x)^2-c\tau^2_x-d(\tau^2_x)^2-e-f-g+\lambda^1_y \tau^1_y
\label{eq:4dqfy1} \\
\tilde{z} & = & -a\tau^1_x-b(\tau^1_x)^2-c\tau^2_x-d(\tau^2_x)^2-e-f-g+\lambda^1_z \tau^1_z \label{eq:4dqfz1}
\end{eqnarray}

where

\begin{eqnarray}
a & = & \frac{1}{2}g \tau^1_{x0}-\frac{1}{2}gx^1+\frac{1}{2}
\label{eq:4dqfdefa} \\
b & = & -\frac{g}{4}  \label{eq:4dqfdefb} \\
c & = & -\frac{1}{2}g \tau^2_{x0}-\frac{1}{2}gx^2+\frac{1}{2}
\label{eq:4dqfdefc} \\
d & = & \frac{g}{4}  \label{eq:4dqfdefd} \\
e & = & \frac{g}{4}\left((x^2)^2-(x^1)^2+(\tau^2_{x0})^2-(\tau^1_{x0})
+2\tau^1_{x0}x^1+2\tau^2_{x0}x^2 \right)  \label{eq:4dqfdefe} \\
f & = & \frac{x^1-x^2}{2}  \label{eq:4dqfdeff} \\
g & = & -\frac{\tau^1_{x0}+\tau^2_{x0}}{2}  \label{eq:4dqfdefg}
\end{eqnarray}

\section{Conclusions}

In this work we have built explicit maps from Cartesian coordinates to emission coordinates, for suitably chosen
set of emitters, starting from the underlying assumption of knowing the whole details of the emitters
world-lines, i.e., for instance the velocities and the positions at a given time. The approach that we have
outlined allows us to express the uncertainties on the Cartesian coordinates in terms of the uncertainties on
the emission coordinates and on the parameters of the world-lines; the definition of the coordinate domain (i.e.
the space-time region where the emission coordinates are unambiguously defined) depends on the emitters
world-lines.

In particular, we have obtained explicit maps in Minkowski 1+1 dimensional space-time, for inertial emitters and
accelerated ones. The latter can be also interpreted as emitters at rest in a gravitational field. The same
procedure has been generalized to 1+3 dimensional space time, both in the inertial case and in the case where a
small "gravitational" field has been introduced. Actually, in order to give a more operational definition of the
emission coordinates systems more general and realistic situations should be studied, nonetheless the toy models
that we have studied here are suitable to suggest the basic ideas of this fully relativistic approach to
positioning systems.

\section*{ACKNOWLEDGMENTS}
The authors would like to thank Dr. D. Bini and Dr. A. Geralico for useful discussions. M.L.R acknowledges
financial support from the Italian Ministry of University and Research (MIUR) under the national program 'Cofin
2005' - \textit{La pulsar doppia e oltre: verso una nuova era della ricerca sulle
pulsar}. \\

\end{document}